\documentclass[11pt,twoside]{article}
\usepackage{asp2010}

\resetcounters

\begin{document}

\title{The Snapshot Hubble U-band Cluster Survey; A cluster complex in NGC\,2146}
\author{Angela Adamo,$^1$ and Linda J. Smith$^2$
\affil{$^1$Max-Planck-Institut for  Astronomy, K\"onigstuhl 17, D-69117 Heidelberg, Germany}
\affil{$^2$Space Telescope Science Institute and European Space Agency, 3700 San Martin Drive, Baltimore, MD 21218, USA}}

\begin{abstract}
We present the Snapshot Hubble $U$-band Cluster Survey ({\it SHUCS}), an ongoing deep $U$-band imaging survey of nearby star-forming galaxies. Thanks to the information provided by the $U$ band, together with archival {\it Hubble Space Telescope} (HST) optical data, we are able to constrain reliable ages, masses, and extinctions of the cluster populations of these galaxies. We show some preliminary results from the study of one of the {\it SHUCS} galaxies, NGC 2146. Using the recovered cluster ages we try to understand the propagation of the star formation in one of the tidal streams where a ring-like cluster complex has been found. The Ruby Ring, so named due to its appearance, shows a clear ring-like distribution of star clusters around a central object. We find evidence of a spatial and temporal correlation between the central cluster  and the clusters in the ring. The Ruby Ring is the product of an intense and localised burst of star formation, similar to the extended cluster complexes observed in M\,51 and the Antennae, but more impressive because is quite isolated. We discuss the formation of the Ruby Ring in a "collect \& collapse" framework. The predictions made by this model agree quite well with the estimated bubble radius and expansion velocity produced by the feedback from the central cluster, making the Ruby Ring an interesting case of triggered star formation.
\end{abstract}

\section{The Snapshot Hubble $U$-band Cluster Survey; {\it SHUCS}}

Local star-forming galaxies host populations of clusters, often with masses and densities that rival globular clusters, objects once thought to form only in the early universe. Despite the progress in understanding cluster formation, evolution, and how they relate to the star formation process as a whole, definitive answers for many questions still remain elusive. The physical properties of clusters retain key information about the nature of the star formation event (i.e. its intensity), the environment, and the physical condition of their hosts at the moment they formed.
What fraction of stars forms in clusters? Which fraction survives? Does this depend on environment? The answer to these questions will help us to understand how star formation proceeds in galaxies. 

{\it SHUCS} is a project aimed at characterising the star cluster populations of 22 nearby galaxies ($\leq$ 22 Mpc). We combine new WFC3 $U$-band imaging with archival HST $BVI$ imaging, to measure the ages, masses and sizes of large samples of young clusters in nearby star-forming galaxies. The aim of this survey is to create a statistically meaningful sample of cluster population to address fundamental questions regarding cluster properties, survival rates, size distributions, formation histories, and environmental dependencies. Our target list covers mostly face-on spiral galaxies, sampling a wide range in mass, size and luminosity, and includes six hosts of active nuclei. An overview of the survey, data reduction and analysis, and goals will soon be released (Konstantopoulos et al., in prep). In this proceeding we will present one-object study focused on the starburst galaxy NGC\,2146 (Adamo et al. in prep) and the analysis of a unique ring-like cluster complex localised at the edge of one of the tidal streams of the galaxy \citep{2012arXiv1205.6204A}.

\begin{figure}
\resizebox{0.53\hsize}{!}{\rotatebox{0}{\includegraphics{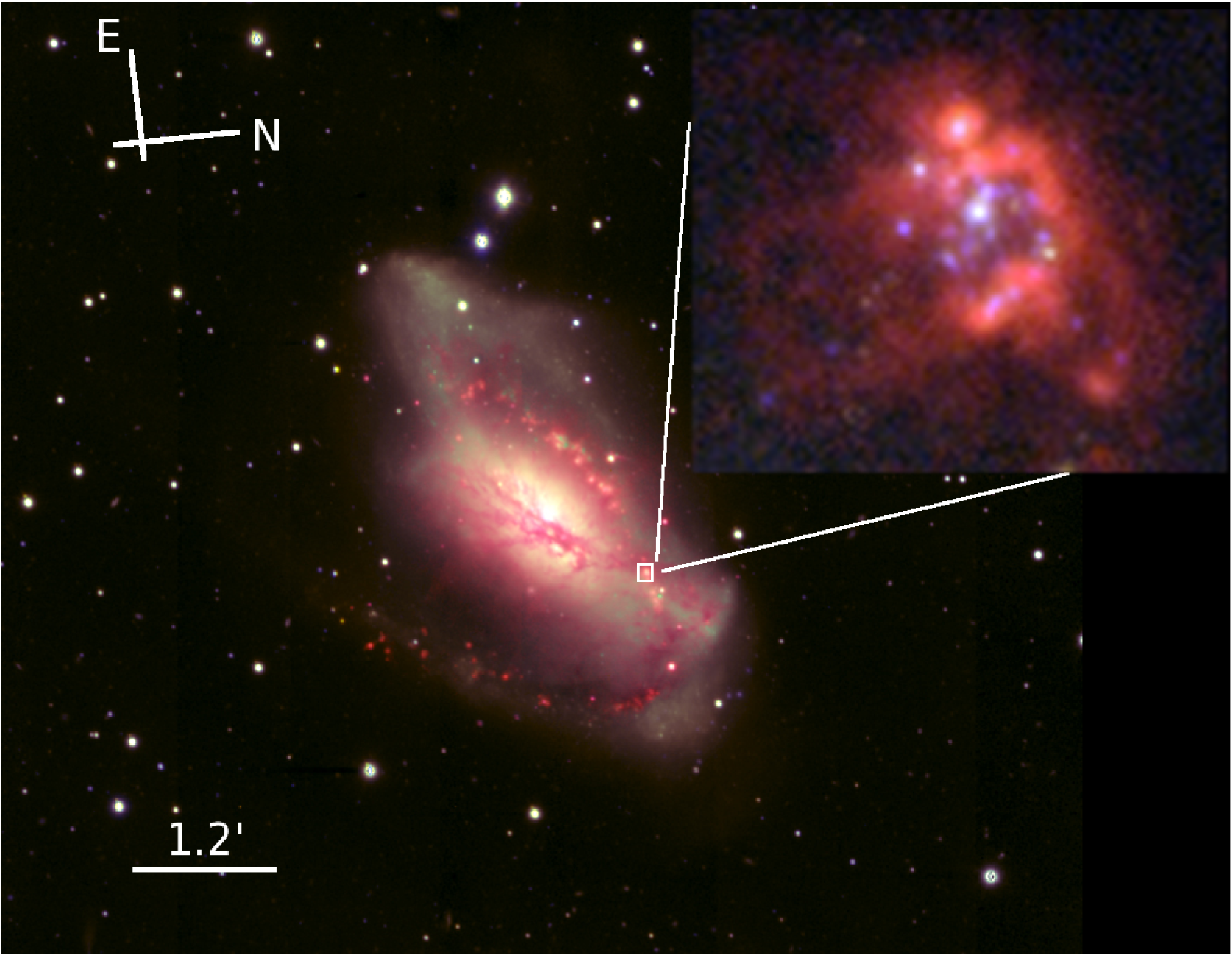}}}
\resizebox{0.49\hsize}{!}{\rotatebox{0}{\includegraphics{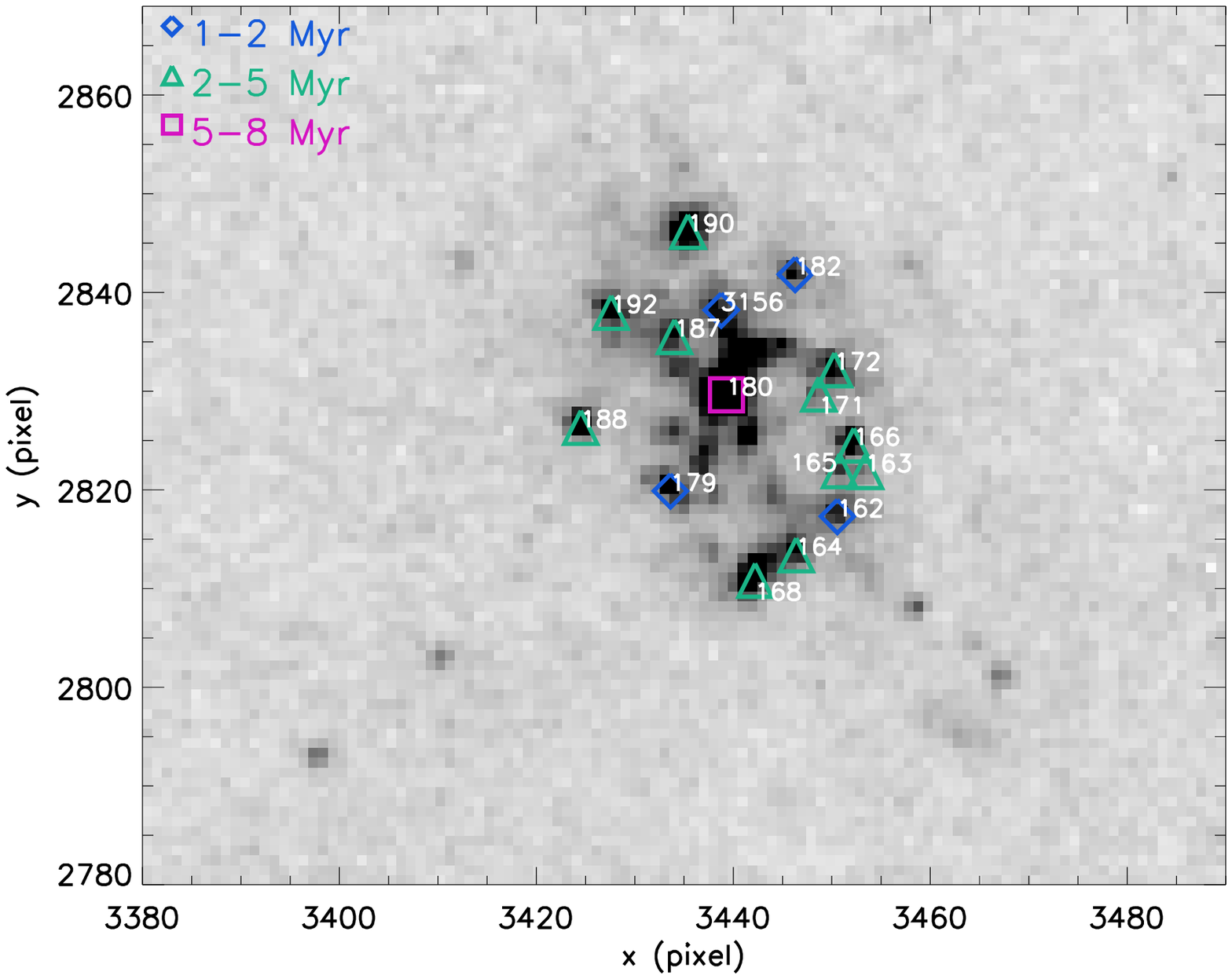}}}
\caption{Left: A color composite of NGC\,2146, using WIYN $B$, $R$, and continuum subtracted H$\alpha$ imaging data in the blue, green, and red channels, respectively. An inset in the upper right corner shows a zoom-in image of the Ruby Ring region, made with {\it HST} data. In the latter, the F658N filter, transmitting H$\alpha$, is in red; the F336W filter, sensitive to the young stellar populations is shown in blue; the F606W and F814W are shown in green and yellow, respectively. The size of the inset is $2.5"$ by $2.4"$ ($\sim130\times125$~pc$^2$ at the assumed distance of 11~Mpc). Right: the cluster ages in color code are overplotted to the F336W image of the ring. Size and orientation are the same as the inset in the right panel. See \cite{2012arXiv1205.6204A} for details on the cluster analysis.}
\label{fig2}
\end{figure}

\section{NGC\,2146; study of a merger system}

The large-scale color composite image of NGC\,2146 (see Figure\ref{fig2}), made of WIYN optical broadband imaging data, directly reveals typical signatures of a merger, e.g tidal streams, intense central starburst. The "merger" nature of the galaxy  is also supported by the extended H{\sc i} reservoir of gas, suggesting an interaction with a low surface brightness galaxy companion \citep{2001A&A...365..360T}. Deep U-band (F336W) imaging of NGC2146 in the SHUCS program has enhanced the accessible archival HST $BVIH_{\alpha}$ data. The WFC3/F336W data has added critical information to the available archival imaging set, allowing us to determine ages, masses, and extinctions of the clusters \citep[see][for details on the data analysis]{2012arXiv1205.6204A}.

\subsection{A ring-like cluster complex in NGC\,2146; properties and formation scenario}

As shown by the inset of the left panel in Figure~\ref{fig2} a ring-like cluster complex is located in one of the tidal streams of NGC\,2146. The appearance of this feature has led us to name it "Ruby Ring". 

\begin{figure}
\resizebox{\hsize}{!}{\rotatebox{0}{\includegraphics{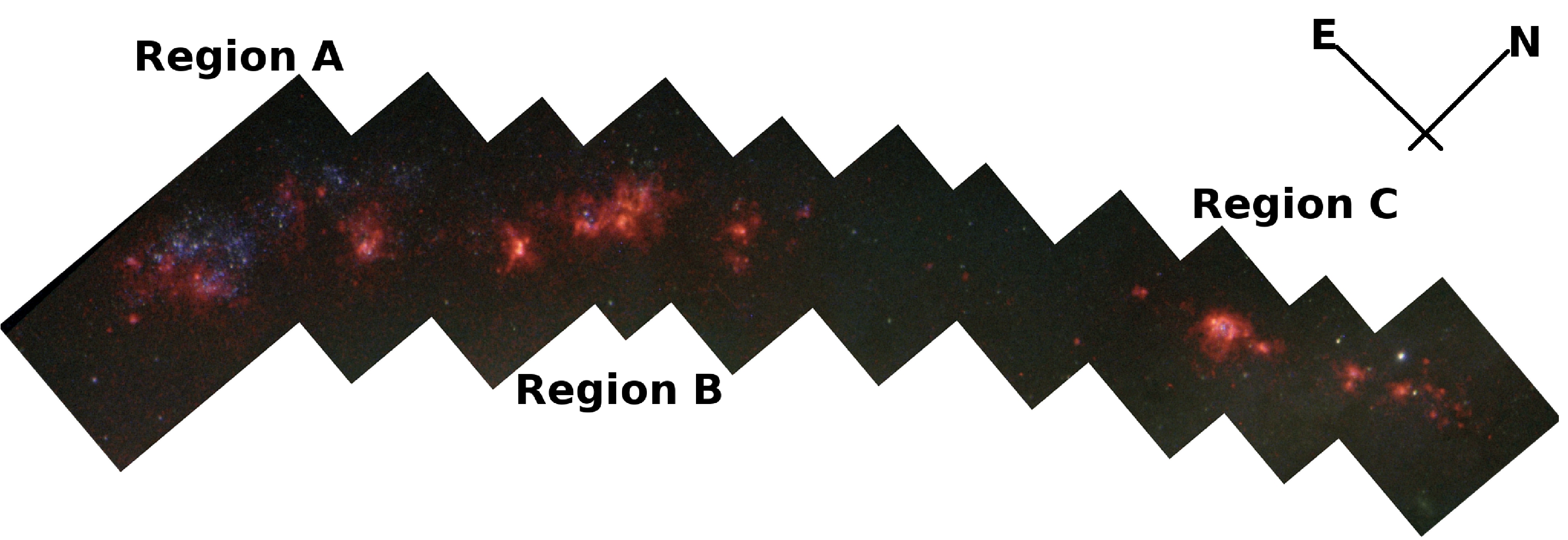}}}\\
\resizebox{0.4\hsize}{!}{\rotatebox{0}{\includegraphics{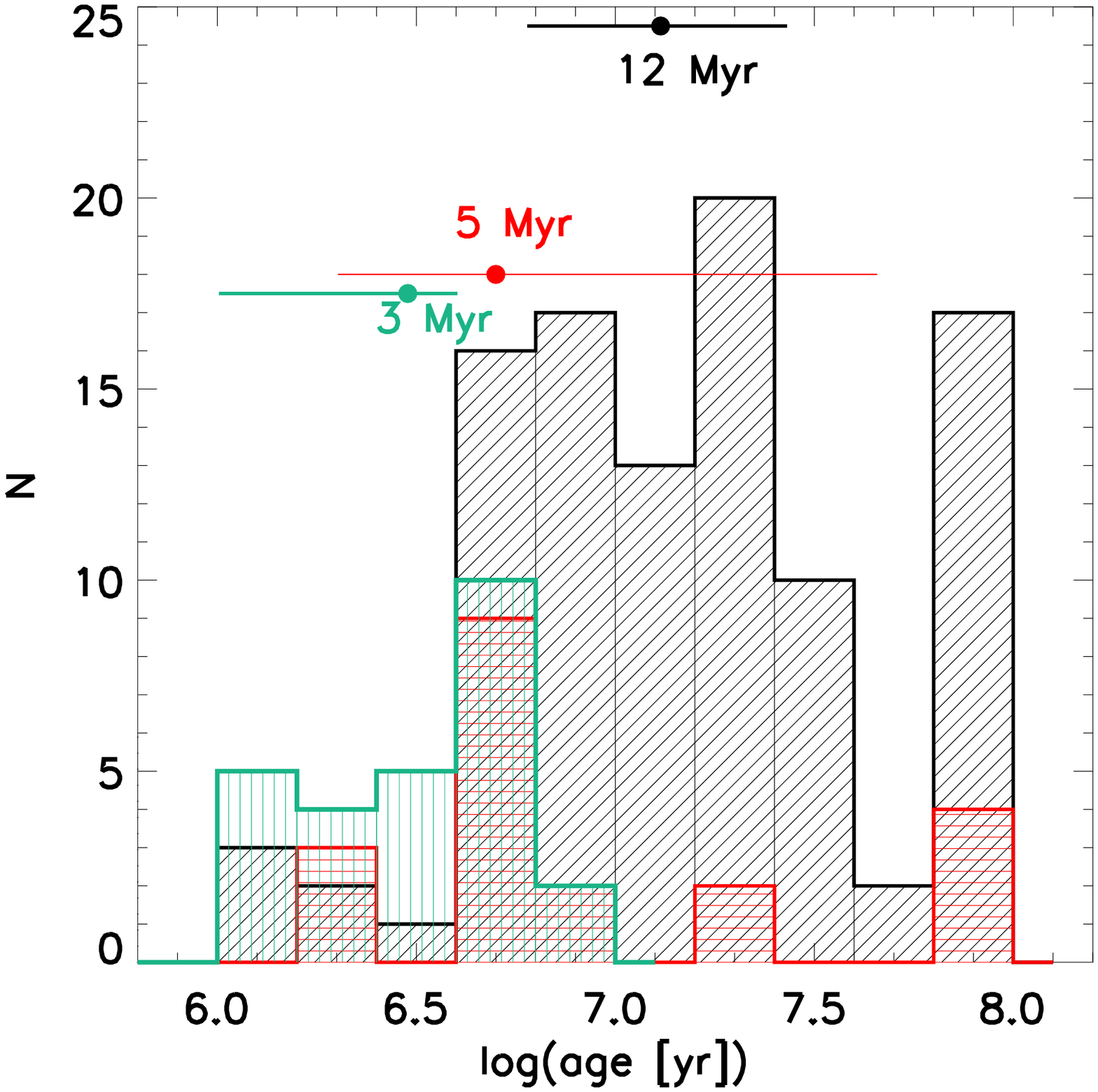}}}
\caption{Upper panel: a color cut-out of the HST composite image of the stream in which the Ruby Ring is located (Region C). The blue and red channels show the F336W and $H_{\alpha}$ filters, while the green and yellow the F606W and F814W frames, respectively. Lower left panel: age distributions of the clusters detected in the 3 regions outlined in the upper panel. For each region we indicated the median and the quartiles (dots and error bars) of the corresponding age histograms. Black has been used for Region A, red for Region B, and green for Region C.}
\label{fig3}
\end{figure}

The recovered cluster ages suggest that star formation in the Ruby Ring's tidal stream started less than 100 Myr ago and has continued until the present (Adamo et al. in prep.). We observe a spatial and temporal correlation in the clusters of the stream (see Figure~\ref{fig3}), e.g. older at the head of the stream (i.e. Region A)  and very young at the tail (Region C). It is not clear yet whether this sequentiality is due to a gravitational instability travelling throughout the streamed gas or to a global instability of the gas in the ejected tail. The Ruby Ring has, thus, formed in a pocket of ejected gas located in a rather low density environment. The shape of the ring suggests that no shear is acting in deforming the system. This is, probably, the most important difference between the Ruby Ring and many galactic and extragalactic systems where triggering has produced arcs or asymmetric bubbles  \citep{2012MNRAS.tmp.2286T, 1998MNRAS.299..643E}. 


An exhaustive description of the cluster analysis of the Ruby Ring can be found in \cite{2012arXiv1205.6204A}, we will discuss here only the main outcomes. We find that the clusters of the Ruby Ring complex have ages younger than 10 Myr. In Figure~\ref{fig2}, right panel, the derived age ranges of the clusters are overplotted on the F336W image of the Ruby Ring, with different colors. The morphology of the ring changes quite drastically from the ultraviolet to H$\alpha$. The $U$-band reveals the sources of the ionising flux, i.e. numerous very young clusters with masses above $10^3$ M$_{\odot}$. The F658N transmits mostly the emission from the photoionized gas surrounding the clusters (red emission in the inset of the left panel in Figure~\ref{fig2}), highlighting the spatially extended H{\sc ii}-regions. The derived best age of the central cluster 180 is of  7.0$^{+86.0}_{-0.0}$ Myr. The mean age of the clusters in the ring is of $\sim 3$ Myr. We derived the total luminosity, L(H$\alpha$), in the region of the Ruby Ring. We found L(H$\alpha$)$=4.6\times 10^{39}$ erg/s, which corresponds to a local star formation rate (SFR) surface density,  $\Sigma_{\textnormal{SFR}}=0.47 $ M$_{\odot}$/yr/kpc$^{2}$. This value is quite similar to the ones found for cluster complexes in other nearby star-forming galaxies like M\,51 and the Antennae \citep{2005A&A...443...79B, 2006A&A...445..471B} and categorise the Ruby Ring as a localised starburst region. The total H$\alpha$ luminosity of the Ruby Ring is $\sim 1/3$ of the L(H$\alpha$) observed in 30 Doradus. 

To explain the formation of the Ruby Ring, we apply a "collect \& collapse" scenario \citep{1977ApJ...214..725E}. The latter predicts that an expanding H{\sc ii} region sweeps up the surrounding remaining low density material in a shell. This shell will eventually fragment and collapse, forming new stars. We tested the predictions made by this model deriving the bubble radius and expansion velocity produced by the feedback from the central cluster.  We found that the location of the ring is quite similar to the determined position of the expanding bubble produced by the central cluster. At the time the star formation was triggered in the ring, the expanding velocity of the shell produced by the "collect \& collapse" model agrees within a factor of 2 with the derived velocity of the central expanding bubble. The "collect \& collapse" model is, thus, able to explain, at the same time, the morphology, the spatial and temporal correlation observed between the central clusters and the clusters in the ring, and the stellar mass ratio between the central source and the triggered ones. Triggered star-forming regions are usually observed in the Milky Way and in local star-forming galaxies. However, what makes the Ruby Ring a unique system is the formation of a well-shaped ring of clusters.

\acknowledgements AA is grateful to the conference organisers to make this celebration possible. AA and LJS thank and acknowledge their collaborators on the {\it SHUCS} project; namely,  I.~S. Konstantopoulos, J.~S. Gallagher, N. Bastian, S.~S. Larsen, E. Silva-Villa, M.~S. Westmoquette, J. Ryon,  E. Zackrisson, J.~C. Charlton, and D.~R. Weisz. We deplore the choice made by Utrecht University board to close the Astronomy Department. It has been a lost not only for the small astronomy world but for the entire society.

\end{document}